\documentclass[12pt]{article}
\usepackage[left=2.5cm,top=2.5cm, bottom=1.2in, right=2.5cm]{geometry}
\usepackage{IEEEtrantools}
\usepackage[cmex10]{amsmath}
\usepackage{cite}
\usepackage{amssymb, mathrsfs, amsfonts, amsthm}
\usepackage{mathtools}
\usepackage[usenames,dvipsnames]{xcolor}
\usepackage{graphicx}
\usepackage{paralist}
\usepackage[colorlinks=true]{hyperref}
\usepackage{caption}
\usepackage{subcaption}
\usepackage{pgfplots}
\usepackage{tikz}
\usepackage{float}
\usetikzlibrary{plotmarks}
\usetikzlibrary{arrows.meta}
\usepgfplotslibrary{patchplots}
\usepackage{grffile}

\theoremstyle{theorem}
\newtheorem{theorem}{Theorem}
\newtheorem{lemma}{Lemma}

\newtheorem{corollary}{Corollary}
\newtheorem{definition}{Definition}

\newtheorem*{example*}{Example}
\newtheorem{remark}{Remark}
\newtheorem*{remark*}{Remark}

\def\iid{\mathop{\mathrm{IID}}}  
\def\crdf{\mathop{\mathrm{CRDF}}}     
\def\cldf{\mathop{\mathrm{CLF}}}     
\def\rvs{\mathop{\mathrm{RVs}}}     
\def\pmf{\mathop{\mathrm{PMF}}}     

\newcommand{\E}[1]{\mathbb{E}\left[#1\right]}
\newcommand{\Eb}[1]{\mathbb{E}\big[#1\big]}
\newcommand{\floor}[1]{\left\lfloor #1 \right\rfloor}

\renewcommand{\H}[1]{{H}\!\left( #1 \right)}
\newcommand{\Hb}[1]{{H}\big( #1 \big)}
\newcommand{\I}[2]{{I}\!\left( #1 ; #2 \right)}
\newcommand{\Ib}[2]{{I}\big( #1 ; #2 \big)}
\newcommand{\ud}{\,\mathrm{d}}

\allowdisplaybreaks
\allowbreak

\begin{document}
\title{Secure Block Joint Source-Channel Coding with Sequential Encoding}

\author{
	Hamid Ghourchian,
	Tobias J. Oechtering, and
	Mikael Skoglund
	\thanks{This is a long version of the accepted ISIT 2023 paper. This work was supported in part by the Swedish Strategic Research Foundation.}
	\thanks{The authors are with the Department of Intelligent Systems, Division of Information Science and Engineering, at KTH Royal Institute of Technology, 10044 Stockholm, Sweden (e-mail: \mbox{\{hamidgh; oech; skoglund\}@kth.se}).}
}

\renewcommand\footnotemark{}
\renewcommand\footnoterule{}

\date{}
\maketitle

\begin{abstract}
We extend the results of Ghourchian et al. \cite{HFTM21} to joint source-channel coding with eavesdropping. 
Our work characterizes the sequential encoding process using the \emph{cumulative rate distribution functions} ($\crdf$) and includes a security constraint using the \emph{cumulative leakage distribution functions} ($\cldf$). 
The information leakage is defined based on the mutual information between the source and the output of the wiretap channel to the eavesdropper. 
We derive inner and outer bounds on the achievable $\crdf$ for a given source and $\cldf$, and show that the bounds are tight when the distribution achieving the capacity of the wiretap channel is the same as the one achieving the capacity of the channel.\\

\end{abstract}

\section{Introduction}
This work deals with a joint source-channel coding problem where an \emph{independent and identically distributed} ($\iid$) source is transmitted through a wiretap channel. 
The decoder provides a lossy reconstruction with an additive distortion constraint while the eavesdropper receives a noisy version of the codeword. 
The eavesdropper sequentially observes the transmitted sub-blocks.
Security constraints prevent the sequential encoding from revealing too much information about the part of the source that has been transmitted so far.
Since the encoding process occurs in sequential sub-blocks, the security constraints are required to avoid information about the source from being revealed early.
Likewise \cite{HFTM21}, the goal is to determine achievable source-channel coding with a specified fidelity and practical applications include communication between vehicles and a central processing node in a sensor network. 
The information leakage constraint, represented by a \emph{cumulative rate leakage function} ($\cldf$; see Definition~\ref{def:CLDF}), is motivated by scenarios where there is an eavesdropper listening to the messages or information should only be revealed according to specific patterns.
Our paper extends the work in \cite{HFTM21} by considering the joint source-channel coding problem in a wiretap channel setting. 

In the classical rate-distortion theory, first introduced in \cite{Shannon59}, the relationship between the achievable distortion and the rate for a non-causal encoder and decoder pair is explored. 
This theory characterizes the fundamental trade-off between the two. 
The rate-distortion function for causal codes was introduced in \cite{Neuhoff82}. 
This variation of classical rate-distortion theory restricts the reconstruction of a current sample to only rely on current and past samples. 
The code stream itself can have variable rates and be non-causal. 
\cite{Weissman05} extended the framework to allow for side information. 
One type of causal source coding is zero-delay coding, where the encoding and decoding are done in real-time. This concept can be seen in \cite{Linder01}. 
Recently, both information theorists and the control community have paid particular attention to the study of causal and zero-delay source coding. 
These types of compression schemes seem to be suitable for determining the fundamental performance limits in closed-loop control systems \cite{Stavrou18, Tanaka18}. 
Causal and zero-delay source coding have also been applied in the context of source-channel coding, or source coding with finite memory \cite{Akyol14, Matloub06, Merhav03}. 
In summary, sequential source coding, as described in works like \cite{Viswanathan00, Ma11}, is a type of source coding that resembles the framework in question, especially when the number of encoders is very high. 
However, this approach aims to understand the rate region for a limited number of encoders, while the current framework not only considers the security constraint but also characterizes the rate profile for an infinite number of encoders.

The concept of secure communication from an information-theoretical perspective was first introduced by Shannon in \cite{Shannon49}. 
Later, Wyner introduced the wiretap channel \cite{Wyner75} and showed that it was possible to send information with weak secrecy if the eavesdropper's channel was a degraded version of the channel from the encoder to the decoder.
 There are two main approaches to secure communication using information-theoretic tools: one where the encoder and decoder agree on a secret key before the transmission of the source and the other where the decoder and eavesdropper have different versions of side information, achieving secrecy through this difference. 
Shannon's work in \cite{Shannon49} showed that if the rate of the secret key is equal to or greater than the entropy of the source, the transmission of a discrete memoryless source is secure. 
Yamamoto in \cite{Yamamoto97} looked at combining secrecy with rate-distortion theory in secure source coding scenarios. 
In \cite{Prabhakaran07}, the authors explored lossless source coding with side information at both the decoder and eavesdropper when there was no rate constraint between the encoder and decoder. 
The authors in \cite{Gunduz08} examined a setup with side information at the encoder and coded side information at the decoder.
In \cite{Villard13, Schieler14}, the problem of secure lossy source coding and the use of secret keys in communication systems to ensure distortion at an adversary were studied from different perspectives. 
In \cite{Kaspi15}, two source coding models were considered with the added constraint of secrecy.

In \cite{HFTM21}, the authors determined the attainable ranges for the variables $\crdf$, $\cldf$, and distortion in the absence of a channel between the sender and receiver or eavesdropper.
This study further improves upon that work by finding inner and outer bounds for the attainable ranges and showing that these bounds coincide under certain conditions where the same distribution maximizes both the secrecy capacity and the channel capacity.
While in \cite{HFTM21} the encoder's transmission could be fully disclosed by the eavesdropper, in this work the transmission goes through a channel, and the amount of information leaked to the eavesdropper is related to the data transmission rate through the channel.

The paper is organized in the following manner. 
In Section \ref{sec:ProblemStatement}, the authors formally introduce their concepts of $\crdf$, $\cldf$, and the sequential joint source-channel coding scheme. 
The main results are presented in Section \ref{sec:main_results}. 
In Section \ref{sec:usefulLemmas}, several lemmas that are used in the main results are given. 
The proofs of the results are given in the Section \ref{sec:proofs}, and the paper concludes in Section \ref{sec:conclusion}.

\emph{Notations}:
Sets and random variables are denoted by calligraphic and capital letters, respectively. Their realizations are denoted by lower case letters.
$\mathbb{N}$, $\mathbb{Q}$, and $\mathbb{R}$ denote the set of integer, rational, and real numbers, respectively.
The probability mass function of a random variable $X$ with realizations $x$ is denoted by $p_X(x)$ or $p(x)$. The conditional probability mass function of $Y$ given $X=x$ is denoted by $p_{Y|X}(y|x)$ or $p(y|x)$.
The sequence $(x_{m}, x_{m+1}, \ldots, x_n)$ is denoted by $x_{m}^{n}$. 
If $m=1$, the notation $x^n$ can be used instead of $x_1^n$.
The expected value of a random variable $X$ is denoted by $\mathbb{E}[X]$.
Logarithms are in base $2$ unless otherwise stated. 
The term ``w.r.t.'' stands for ``with respect to''.

\section{Problem Statement and New Definitions} \label{sec:ProblemStatement} 
\begin{figure}[h]
	\centering
	\includegraphics[trim = 0 90 0 80, clip, width=\columnwidth, scale=1]{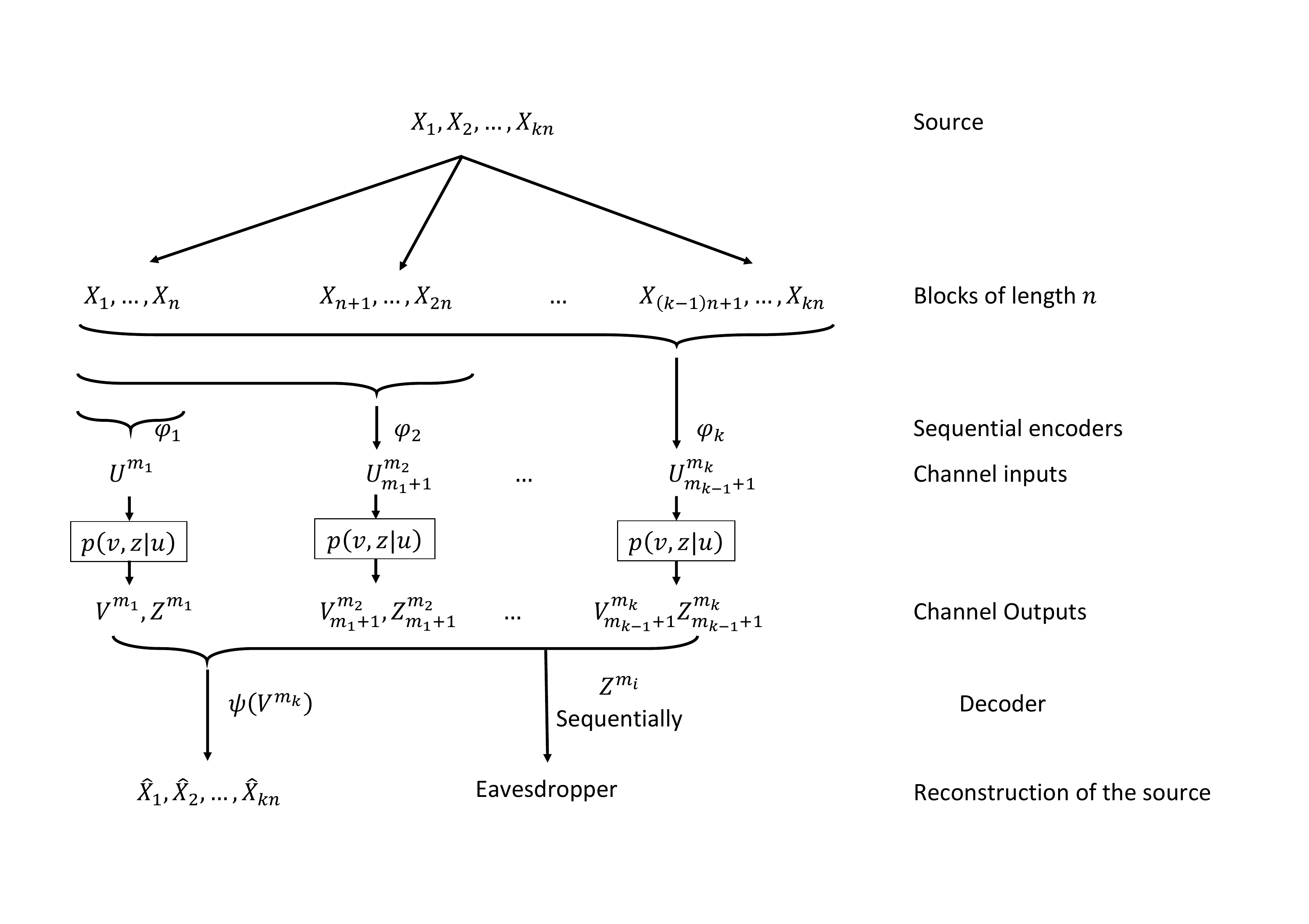}
	\caption{The encoders $\varphi_1,\ldots,\varphi_k$ encode the current and past sub-blocks of length $n$ in sequence. The rate profile is determined by the rate of these encoders. The decoder $\psi$ decodes all $V^{m_k}$ channel outputs at the same time, while the eavesdropper sees the $Z^{m_i}$ channel outputs as they are transmitted sequentially.}
	\label{fig:system}
\end{figure}

In this section, we provide our problem formulation and some new definitions.
As shown in Fig.~\ref{fig:system}, the source block consists of $nk$ $\iid$ random variables that are divided into $k$ sub-blocks of length $n$.
For each sub-block, there is an encoder that has access to the source symbols of the previous and current sub-blocks, but not the future sub-blocks, resulting in causal encoding.
The output of each encoder, $U_{m_{i-1}+1}^{m_i}$, passes through a wiretap channel with conditional probability mass function $p(v,z|u)$. 
Note that the decoder is not restricted to be causal, meaning it can wait until it receives the outputs of all $V^{m_k}$ channels for the entire source sequence before making a decision.

The definition of \emph{regular function}, $\crdf$, and $\cldf$ are the same as \cite[Definitions 1-3]{HFTM21}, while, their interpretations are different.
Nevertheless, we express them again to make the paper self-contained.

The encoding of the sub-blocks follows a particular rate-distortion function ($\crdf$). 
Additionally, a constraint on the leakage of information ($\cldf$) is imposed due to the possibility of an eavesdropper intercepting the encoded sub-block messages. 
This leakage constraint is expressed as the gradual accumulation of mutual information over time.

\begin{definition}[Regular Cumulative Function] \label{def:Regular Cumulative Function}
	A function $F \colon [0,1] \to [0,\infty)$ is a \emph{regular cumulative function} if it satisfies the following properties:
	\begin{enumerate}
		\item
		Cumulation: $F$ is non-decreasing,
		\item
		Zero initial value: $F(0) = 0$,
		\item
		Regularization: $F$ is continuous from the right, i.e.,
		$\lim_{\beta\searrow\alpha}F(\beta) = F(\alpha)$ for all $\alpha\in [0,1)$.
	\end{enumerate}
\end{definition}
Regular cumulative functions represent both the cumulative rate at which the encoding is allowed and the allowable leakage.

\begin{definition}[Cumulative Rate Distribution Function] \label{def:CRDF}
	A function $G \colon [0,1] \to [0,\infty)$ is a $\crdf$ if $G$ is a \emph{regular cumulative function} characterizing the cumulative rate at which encoding is allowed.
	The domain of the function $G$ represents the normalized time of blocks for the transmission of the whole sequence of blocks; as a result, if there are $k$ blocks, $G(\alpha)$ represents the accumulated rate until block $\floor{\alpha k}$ ends.
\end{definition}

Because the function characterizes the cumulative rate, it is non-decreasing.
There is no need to consider an available rate before the start of the sequence; so, it has zero initial value.

\begin{definition}[Cumulative Leakage Function] \label{def:CLDF}
	A function $L \colon [0,1] \to [0,\infty)$ is a $\cldf$ if $L$ is a \emph{regular cumulative function} characterizing the cumulative leakage constraint over time.
	The domain of the function $L$ represents the normalized time of blocks w.r.t. the whole time of the sequence; as a result, if there are $k$ blocks of $n$ symbols, $L(\alpha)$ represents the allowable leakage of the compressed messages until block $\floor{\alpha k}$ ends about the $n \floor{\alpha k}$ first symbols of the source, i.e., the leakage of $U^{m_{\floor{\alpha k}}}$ about $X^{n\floor{\alpha k}}$.
\end{definition}

The assumption is that the amount of information leaked does not decrease as time passes. 
This aligns with the definition of mutual information as a measure of leakage because $\I{X^{ni}}{U^{m_i}}\geq \Ib{X^{n(i-1)}}{U^{m_{i-1}}}$ (note that $m_i\geq m_{i-1}$).
As a result, the leakage function is non-decreasing and has a zero initial value. 
For a fixed number of sub-blocks $k$, as the number of samples $n$ increases, the values of the rate and leakage profiles at certain points determine the overall rate and leakage profiles. 
However, for larger values of $k$, more points of the rate and leakage functions become necessary. 
Thus, for a fixed $k$, evaluating the rate and leakage profiles is equivalent to using step-wise functions based on the values at certain points.

\begin{definition}[Sequential encoding with $(G, k, n)$- source-channel codes] \label{chp:SeqJSC:def:Code}
	\sloppy Assume $(X_1, X_2, \ldots, X_{n k})$ is a block of $\rvs$ each one defined on domain $\mathcal{X}$.
	A $(G, k, n)$-source-channel code, where $(k, n)\in\mathbb{N}$, and $G \colon [0,1] \to [0,\infty)$ is a $\crdf$ according to Definition \ref{def:CRDF}, consists of
	\begin{itemize}
		\item an ensemble of $k$ sequential encoders $\varphi_1,\ldots,\varphi_k$ such that each one of them assigns a sequence of the inputs of the channel to the source sequence blocks received so far, i.e., for each $i\in\{1,\ldots,k\}$,
		\begin{IEEEeqnarray*}{rCl}
			\varphi_i &\colon& \mathcal{X}^{i n} \to \mathcal{U}_{m_{i-1}+1}^{m_i}, \\
			&& x^{i n} \mapsto u_{m_{i-1}+1}^{m_i},
			\quad m_i := \floor{n k r_i} + m_{i-1}, 
			\;\; m_0 = 0,
		\end{IEEEeqnarray*}
		where
		\begin{equation} \label{chp:SeqJSC:eqn:def R}
			r_i = G\left( \tfrac{i}{k} \right) - G\left( \tfrac{i-1}{k}\right),
			\qquad i=1,\ldots,k,
		\end{equation}
		\item a decoder, $\psi$, that reconstructs $\hat{x}^{n k}$ based on the output of the channels, i.e.,
		\begin{IEEEeqnarray*}{rCl}
			\psi &\colon& \mathcal{V}^{m_k} \to \hat{\mathcal{X}}^{n k}, \\
			&& v^{m_k} \mapsto \hat{x}^{n k},
		\end{IEEEeqnarray*}
		where $\hat{\mathcal{X}}$ denotes the reconstruction domain.
	\end{itemize}
\end{definition}

Next, we introduce the definition of the achievable set of pair of $\crdf$s and $\cldf$s.
\begin{definition}[Achievable $\cldf$-secure $\crdf$ for lossy compression] \label{chp:SeqJSC:def:Achievable R(D)}
	Assume that $X_1, X_2, \ldots$ is a sequence of $\rvs$, each with support $\mathcal{X}$ and $p(v|u)$ is a discrete memoryless channel.
	A $\crdf$ $G$ is said to be achievable to encode the sequence $X_1, X_2, \ldots$, given $\cldf$ $L$ and an average expected distortion level less than $\bar{d}$, if for any $k\in\mathbb{N}$, there exists a sequence of $(G, k, n)$-source codes, for $n\in\mathbb{N}$, with output support $\hat{\mathcal{X}}$, such that
	\begin{IEEEeqnarray}{l}
		\limsup_{n\to\infty}{\E{d(X^{n k},\hat{X}^{n k})}}
		\leq \bar{d}, \label{chp:SeqJSC:eqn:limsupEd<bard}\\
		\frac{1}{n k}\Ib{X^{i n}}{Z^{m_i}} \leq L\left(\tfrac{i}{k}\right),
		\qquad \forall i\in\{1, 2, \ldots, k\}, \label{chp:SeqJSC:eqn:mainI<L}
	\end{IEEEeqnarray}
	where for the given distortion function $d \colon \mathcal{X}\times\hat{\mathcal{X}} \to [0,\infty]$, we have
	\begin{equation*}
		d(X^{n k},\hat{X}^{n k})
		:= \frac{1}{n k} \sum_{i=1}^{n k}{d(X_i,\hat{X}_i)}.
	\end{equation*}
	Expression \eqref{chp:SeqJSC:eqn:mainI<L} denotes the normalized amount of leakage of the first $i$ blocks, while the leakage of the messages until the end of block $i$ has been measured.
\end{definition}

\section{Main Results} \label{sec:main_results}
In this section, we propose inner and outer bounds to characterize the set of achievable pairs of $\crdf$ and $\cldf$.

The given sequence $X_1, X_2, \ldots$ is considered to be $\iid$ with a $\pmf$ $p(x)$ and a finite set of possible values, known as the support, denoted as $\mathcal{X}$. 
Additionally, a distortion function $d$ that maps values from $\mathcal{X}\times\hat{\mathcal{X}}$ to a range of $[0,\infty]$ is assumed. 
The distortion-rate function $D$ (as defined in Definition \ref{def:classical R(D)}) is also bounded, meaning the maximum value of $D(R)$ is finite for all $R\geq 0$. 
The capacity of the channel with conditional probability mass function $p(v|u)$, $C$, and the secrecy capacity of the wiretap channel with conditional probability mass function $p(v,z|u)$, $C_\mathrm{WT}$, are both bounded:
\begin{IEEEeqnarray}{rCl}
	C &:=& \max_{p(u)}\I{U}{V}, \label{eqn:ChannelC}\\
	C_\mathrm{WT} &:=& \max_{p(w,u)} \I{W}{V} - \I{W}{Z}.
\end{IEEEeqnarray}

First we state the outer bound.
\begin{theorem} \label{CHP:SEQJSC:THM:R(D):CONV}
	For given $\crdf$ $G$ and $\cldf$ $L$, define $G^\mathrm{eff}_\mathrm{out} \colon [0,1] \to [0,\infty)$ as
	\begin{IEEEeqnarray}{l} 
		G^\mathrm{eff}_\mathrm{out}(\alpha) \nonumber\\
		\qquad:= \max\Big\{C_\mathrm{WT} G(\alpha), \nonumber\\
		\qquad\qquad C G(\alpha) - \sup_{\beta\in[0,1]}\left((C-C_\mathrm{WT}) G(\beta) - L(\beta)\right)\Big\}. \IEEEeqnarraynumspace\label{chp:SeqJSC:eqn:GL}
	\end{IEEEeqnarray}
	If the $\crdf$ $G$, for a given $\cldf$ $L$, is achievable, with distortion level $\bar{d}$, in the sense of Definition \ref{chp:SeqJSC:def:Achievable R(D)}, We have
	\begin{equation} \label{chp:SeqJSC:eqn:intDG<d}
		\int_0^1 {D\left(\frac{\ud \hat{G}_\mathrm{out}^\mathrm{eff}}{\ud \alpha}(\alpha) \right) \ud \alpha}
		\leq \bar{d},
	\end{equation}
	where $\hat{G}^\mathrm{eff}_\mathrm{out} \colon [0,1] \to [0,\infty)$ is the envelope of the function $G_\mathrm{out}^\mathrm{eff}$ in the sense of Definition \ref{def:ConvHull} and $D(\cdot)$ is the distortion-rate function.
\end{theorem}

\begin{IEEEproof}
	See Section \ref{chp:SeqJSC:subsec:prf:R(D):conv}.
\end{IEEEproof}

Next, we state the inner bound.
\begin{theorem} \label{CHP:SEQJSC:THM:R(D):ACH}
	Define $G^\mathrm{eff}_\mathrm{in} \colon [0,1] \to [0,\infty)$ as
	\begin{IEEEeqnarray}{l} 
		G^\mathrm{eff}_\mathrm{in}(\alpha) \nonumber\\
		\quad :=
		\max\Big\{C_1 G(\alpha), \nonumber\\
		\quad\qquad C_2 G(\alpha) - \sup_{\beta\in[0,1]}\left((C_2-C_1) G(\beta) - \frac{L(\beta)}{\ell}\right)\Big\}, \IEEEeqnarraynumspace\label{chp:SeqJSC:eqn:GL}
	\end{IEEEeqnarray}
	for some $C_1 < C_2$ and $\ell > 0$ such that the set of rate-leakage pairs 
	\begin{equation*}
		\left\{(R, R_L) \colon R_L > \max\{0, \ell (R-C_1)\}, R < C_2\right\}
	\end{equation*}
	is achievable for the wiretap channel $p(v,z|u)$ (see Theorem \ref{thm:bg:wt:RRL}).
	Then, the $\crdf$ $G$ is achievable, given $\cldf$ $L$, with distortion level $\bar{d}$, in the sense of Definition \ref{chp:SeqJSC:def:Achievable R(D)}, if
	\begin{equation} \label{chp:SeqJSC:eqn:intDG<d}
		\int_0^1 {D\left(\frac{\ud \hat{G}_\mathrm{in}^\mathrm{eff}}{\ud \alpha}(\alpha) \right) \ud \alpha}
		\leq \bar{d},
	\end{equation}
	where $\hat{G}^\mathrm{eff}_\mathrm{in} \colon [0,1] \to [0,\infty)$ is the envelope of the function $G_\mathrm{in}^\mathrm{eff}$ in the sense of Definition \ref{def:ConvHull} and $D(\cdot)$ is the distortion-rate function.
\end{theorem}

\begin{IEEEproof} 	
	See Section \ref{chp:SeqJSC:subsec:prf:R(D):ach}.
\end{IEEEproof}

Compared to \cite{HFTM21}, the key difference here is that the amount of information leaked to the eavesdropper depends on the data transmission rate, whereas in \cite{HFTM21} with no channel present, the leaked information is equal to the amount transmitted by the encoder.
Additionally, the leakage is not necessarily linearly proportional to the transmission rate.
Our proposed inner and outer bounds identify the solution where the leakage increases linearly with the transmission rate beyond a certain point, which is the intersection of two linear diagrams.
The next corollary illustrates a scenario where the rate-leakage diagram has the same type of bounds found in this study, implying that the bounds are tight.
\begin{corollary}
	For wiretap channel $p(v,z|u)$ such that the optimal $\pmf$ $p^*(u)$ in \eqref{eqn:ChannelC} and Theorem \ref{thm:bg:wt:sc} are equal, the bounds are tight.
\end{corollary}

\begin{IEEEproof}
	In this case, from Theorem \ref{thm:bg:wt:RRL}, we obtain that the line from $C_\mathrm{WT}$ to $C$, in the achievable rate-leakage region of the wiretap channel, has slope $1$; as a results, $\ell=1$ and the bounds are tight.
\end{IEEEproof}

\section{Useful Definitions and Lemmas} \label{sec:usefulLemmas}
In this section, we outline several commonly known definitions used in this work and provide proofs for several lemmas used in the demonstration of the main results.

\begin{definition}[Concave-hull or envelope of a function] \cite[p. 119]{Boyd04} \label{def:ConvHull}
	Let $f\colon \mathcal{A} \to \mathbb{R}$ be a function with a convex domain $\mathcal{A}$.
	Then, $\hat{f}\colon \mathcal{A} \to \mathbb{R}$ is the \emph{concave hull or envelope} of $f$ if $\hat{f}$ is a concave function such that $f(x) \leq \hat{f}(x),~\forall x\in \mathcal{A}$, and for any concave function $g \colon \mathcal{A} \to \mathbb{R}$ such that $f(x) \leq g(x),~\forall x\in \mathcal{A}$, we have $\hat{f}(x) \leq g(x),~\forall x\in \mathcal{A}$.
\end{definition}

\begin{definition}[Rate-distortion and distortion-rate functions] \cite[p. 307]{Cover06} \label{def:classical R(D)}
	For a given distortion function $d \colon \mathcal{X}\times\hat{\mathcal{X}} \to [0,\infty)$ and a probability distribution $X\sim p(x)$, the rate distortion function, $R(D)$, is defined based on \cite[p. 307]{Cover06}.
	The inverse of $R(D)$ is the distortion-rate function, denoted by $D(R)$, such that $D(R) = \min\{D \colon R(D) = R\}$.
\end{definition}

The $R(D)$ and $D(R)$ satisfy well-known properties (see, for example \cite{berger:1971,Cover06}), some of which are stated in the next remark.
\begin{remark} \label{cor:R(D)convex}
	$R(D)$ and $D(R)$ are non-increasing and convex functions of $D\in[0,\infty)$ and $R\in[0,\infty)$, respectively. 
	Besides, $R(D)$ and $D(R)$ are continuous w.r.t. $D\in(0,\infty)$ and $R\in(0,\infty)$, respectively.
	Further, if $R(0)<\infty$ or $D(0)<\infty$, then it is continuous w.r.t. $D\in[0,\infty)$ or $R\in[0,\infty)$, respectively.
\end{remark}

The next theorem evaluates the \emph{secrecy capacity} of a wiretap channel.
\begin{theorem} \cite[p. 550]{ElGamal11} \label{thm:bg:wt:sc}
	In the $\mathrm{DM\text{-}WTC}$ $p(y,z|x)$ we have
	\begin{equation*}
		C_\mathrm{WT} = \max_{p(u,x)} \I{U}{Y} - \I{U}{Z},
	\end{equation*}
	where
	\begin{equation*}
		C_\mathrm{WT} := \max\{R\colon (R, 0) \text{ is achievable}\}
	\end{equation*}
	is the \emph{secrecy capacity} of the wiretap channel.
\end{theorem}

The rate-leakage region of a wiretap channel is defined as follows.
\begin{theorem} \cite[Theorem 22.2]{ElGamal11} \label{thm:bg:wt:RRL}
	The rate-leakage region for the $\mathrm{DM\text{-}WTC}$ $p(y,z|x)$ is the set of pair $(R, R_L)$ such that
	\begin{IEEEeqnarray*}{l}
		R \leq \I{U}{Y}, \\
		R \leq \max\left\{0, \I{U}{Y} - \I{U}{Z}\right\} + R_L,
	\end{IEEEeqnarray*}
	for some $\pmf$ $p(u,x)$.
\end{theorem}

\begin{lemma} \label{chp:SeqJSC:lmm:DWTC}
	For a wiretap channel with conditional $\pmf$ $p(z,v|u)$, if the rate-leakage pair $(R,R_L)$ is achievable (see \cite[p. 550]{ElGamal11}), then, it is also achievable for non-uniform message $M$ distributed over the set $\{1,\ldots,2^{nR}\}$.
\end{lemma}

\begin{IEEEproof}
	We utilize pre-coding as following. 
	Let $Q \sim \mathrm{Unif}\{1,\ldots,2^{nR}\}$ independent of $M$.
	Define $M' \equiv M + Q \mod 2^{nR}$; as a result $M'$ is uniformly distributed over $\{1,\ldots, 2^{nR}\}$.
	We know that for $M'$ there exists a sequence of encoders and decoders such that $(R, R_L)$ is achievable, i.e.,
	\begin{IEEEeqnarray}{l}
		\lim_{n\to\infty}{\Pr\{M' \neq \hat{M}'\}} = 0, \label{chp:SeqJSC:eqn:lmm:clas:1}\\
		\limsup_{n\to\infty}{\frac{1}{n}\I{M'}{Z^n}} \leq R_L \label{chp:SeqJSC:eqn:lmm:clas:2}.
	\end{IEEEeqnarray}
	Further, we have
	\begin{IEEEeqnarray}{rCl}
		\I{M'}{Z^n}
		&\stackrel{(a)}{=}& \I{M', Q}{Z^n} \nonumber\\
		&\stackrel{(b)}{=}& \I{M', Q, M}{Z^n} \nonumber\\
		&\stackrel{(c)}{=}& \I{Q, M}{Z^n} \nonumber\\
		&\geq& \I{M}{Z^n \vert Q}, \label{chp:SeqJSC:eqn:lmm:clas:3} 
	\end{IEEEeqnarray}
	where $(a)$ follows from the Markov chain $Q \to M' \to Z^n$;
	$(b)$ follows because $M \equiv M' - Q \mod 2^{nR}$ is a function of $Q$ and $M'$;
	and $(c)$ follows because $M'$ is a function of $Q$ and $M$.
	Hence, from \eqref{chp:SeqJSC:eqn:lmm:clas:2} and \eqref{chp:SeqJSC:eqn:lmm:clas:3}, we obtain
	\begin{IEEEeqnarray}{l}
		\limsup_{n\to\infty}{\frac{1}{n}\I{M}{Z^n \vert Q}} \leq R_L. \label{chp:SeqJSC:eqn:lmm:clas:5}
	\end{IEEEeqnarray} \\
	Assuming $\hat{M} \equiv \hat{M}' - Q \mod 2^{nR}$, we obtain that given $Q$,
	\begin{IEEEeqnarray}{rCl}
		\IEEEeqnarraymulticol{3}{l}{\Pr\{M \neq \hat{M} \mid Q\}} \\
		\qquad&=& \Pr\{M \neq \hat{M}' - Q \mod 2^{nR} \mid Q\} \nonumber\\
		&=& \Pr\{M + Q \mod 2^{nR} \neq \hat{M}' \mod 2^{nR} \mid Q\} \nonumber\\
		&=& \Pr\{M' \neq \hat{M}' \mid Q\}. \label{chp:SeqJSC:eqn:lmm:clas:51}
	\end{IEEEeqnarray}
	Hence, from \eqref{chp:SeqJSC:eqn:lmm:clas:1} and \eqref{chp:SeqJSC:eqn:lmm:clas:51}, we obtain that for any given $Q=q$,
	\begin{IEEEeqnarray}{l}
		\lim_{n\to\infty}{\Pr\{M \neq \hat{M} \mid Q=q \}} = 0. \label{chp:SeqJSC:eqn:lmm:clas:4}
	\end{IEEEeqnarray}
	Therefore, by assuming that $Q$ is known publicly, $(R, R_L)$ is achievable. 
	Now, we are going to find a realization $q$ such that $(R,R_L)$ remains achievable.
	Note that it is valid because $Q$ is independent of $M$ and by taking $Q=q$, the distribution of $M$ does not change. 
	To this end, for any $\epsilon > 0$, we define $E$ as a function of $Q$ such that
	\begin{equation} \label{chp:SeqJSC:eqn:lmm:clas:10}
		E :=
		\begin{cases}
			0 & \text{if }\Pr\{M = \hat{M} \mid Q\} \geq 1-\epsilon, \\
			1 & \mathrm{otherwise}.
		\end{cases}
	\end{equation}
	Hence, from \eqref{chp:SeqJSC:eqn:lmm:clas:4} we obtain that $\Pr\{E = 1\} \to 0$ as $n\to\infty$.
	\begin{IEEEeqnarray*}{rCl}
		\IEEEeqnarraymulticol{3}{l}{\I{M}{Z^n \vert Q}} \\
		\qquad&=& \I{M}{Z^n \vert Q, E} \\
		&\geq& \I{M}{Z^n \vert Q, E=0} \Pr\{E = 0\} \\
		&\stackrel{(a)}{=}& \Pr\{E = 0\} \times \\
		&& \sum_{q} \Pr\{Q=q \mid E=0\} \I{M}{Z^n \vert Q=q},
	\end{IEEEeqnarray*}
	where $(a)$ follows because $E$ is a function of $Q$.
	Hence, there exists $q$ such that 
	\begin{IEEEeqnarray}{l}
		\I{M}{Z^n \mid Q=q} \leq \I{M}{Z^n \vert Q} / \Pr\{E = 0\} \nonumber\\
		\qquad\qquad\stackrel{(a)}{\Rightarrow}\limsup_{n\to\infty}\frac{1}{n}\I{M}{Z^n \mid Q=q} \leq R_L, \\
		E = 0 
		\stackrel{(b)}{\Rightarrow} \lim_{n\to\infty}\Pr\left\{M \neq \hat{M} \mid Q=q\right\} < \epsilon, \label{chp:SeqJSC:eqn:lmm:clas:11}
	\end{IEEEeqnarray}
	where $(a)$ follows from \eqref{chp:SeqJSC:eqn:lmm:clas:5};
	and $(b)$ follows from \eqref{chp:SeqJSC:eqn:lmm:clas:10}.
	Because \eqref{chp:SeqJSC:eqn:lmm:clas:11} is valid for any $\epsilon > 0$, it is also valid for $\epsilon=0$.
	Thus, $(R, R_L)$ is achievable for non-uniform $M$ and the lemma is proved.
\end{IEEEproof}

\begin{lemma} \label{chp:SeqJSC:lmm:CRDFMaj}
	Let $G_1$ and $G_2$ be two $\crdf$s such that the following conditions hold:
	\begin{equation*}
		\begin{cases}
			G_1(\alpha) \geq G_2 (\alpha),
			\qquad \alpha\in[0,1), \\
			G_1(1) = G_2(1).
		\end{cases}
	\end{equation*}
	Then, for $k, n\in\mathbb{N}$ and a sequence of codes $(G_1, k, n)$-code$\colon x^{n k} \mapsto u_{(1)}^{m^k} \mapsto \hat{x}_{(1)}^{n k}$, there exists a sequence of codes $(G_2, k, n)$-code$\colon x^{n k} \mapsto u_{(1)}^{m^k} \mapsto \hat{x}_{(2)}^{n k}$ such that, for any $k\in\mathbb{N}$ and large enough $n$, $\hat{x}_{(1)}^{n k} = \hat{x}_{(2)}^{n k}$, for all $x^{n k}\in\mathcal{X}^{n k}$.
\end{lemma}

\begin{IEEEproof}
The proof is similar to the proof of \cite[Lemma 2]{HFTM21} and follows the same steps.
\end{IEEEproof}

\section{Proofs} \label{sec:proofs}
\subsection{Proof of Theorem \ref{CHP:SEQJSC:THM:R(D):CONV}} \label{chp:SeqJSC:subsec:prf:R(D):conv}
For the distortion, we use the following notation:
\begin{equation} \label{chp:SeqJSC:eqn:partiald}
	d(x_a^b, \hat{x}_a^b)
	:= \frac{1}{b-a+1}\sum_{i=a}^b{d(x_i,\hat{x}_i)}.
\end{equation}

The proof is divided in the following four steps.
\begin{enumerate}	
	\item\label{chp:SeqJSC:itm:ClasConv}
	In this step, we find a relation between the sequence $\tilde{R}_i$, which characterizes the minimum rate needed to satisfy the distortion constraint of the corresponding block, and $r_i$.
	We do not need to send all $\tilde{R}_i$ until the end of block $i$. 
	If we send only a part of it, because decoding is done after all blocks and encoders have access to all past source symbols, then we transfer the rates of the remaining part to the next blocks.
	As a result, the channel usage per source symbol ratio $r_i$ is used to send both a part of $\tilde{R}_i$ and the rates transfered from the previous blocks.
	This strategy will help to satisfy the leakage constraint.\\
	Formally, for any $k\in\mathbb{N}$, there exists a sequence $\tilde{R}_i$, for $i=1,\ldots,k$, such that
	\begin{IEEEeqnarray}{l}
		\sum_{i=j+1}^k{\tilde{R}_i}
		\leq C \sum_{i=j+1}^k{r_i},
		\quad j=0,\ldots,k-1, \label{chp:SeqJSC:eqn:G-Gj/k>sumRhat} \\
		\sum_{i=\ell+1}^k{\tilde{R}_i} 
		\leq L\left(\tfrac{j}{k}\right) 
		+ C_\mathrm{WT}\sum_{i=\ell+1}^j{r_i}
		+ C \sum_{i=j+1}^k {r_i}, \nonumber\\
		\qquad\qquad \ell=0,\ldots,k-1, \quad j=\ell,\ldots,k \label{chp:SeqJSC:eqn:L>sumRhat} \\
		\frac{1}{k} \sum_{i=1}^k {D(k \tilde{R}_i)} \leq \bar{d}. \label{chp:SeqJSC:eqn:avgDhatR<d}
	\end{IEEEeqnarray}
	
	\item\label{chp:SeqJSC:itm:decR}
	In this step, we find a relation between $\tilde{R}_i'$ and $G^\mathrm{eff}_k$ (which is defined later).
	The sequence $\tilde{R}'_i$ is generated from $\tilde{R}_i$ by increasing $\tilde{R}_1$ such that the sequence has the same total rate as the total effective rate of the sequence $R_i$ and by decreasingly sorting the rates $\tilde{R}_i$.\\
	Formally, for any $k\in\mathbb{N}$, there exists a sequence $\tilde{R}'_i$, for $i=1,\ldots,k$, such that
	\begin{IEEEeqnarray}{l}
		\tilde{R}'_1\geq\cdots\geq \tilde{R}'_k, \label{chp:SeqJSC:eqn:R1>Rn} \\
		\sum_{i=1}^j{\tilde{R}'_i}
		\geq G^\mathrm{eff}_k\left(\tfrac{j}{k}\right),
		\qquad j=1,\ldots,k-1, \label{chp:SeqJSC:eqn:sumRt''>GL} \\
		\sum_{i=1}^k{\tilde{R}'_i}
		= G^\mathrm{eff}_k(1), \label{chp:SeqJSC:eqn:totRt''=GL1-GL0} \\
		\frac{1}{k} \sum_{i=1}^k {D(k \tilde{R}'_i)}
		\leq \bar{d}, \label{chp:SeqJSC:eqn:avgDhatR''<d}
	\end{IEEEeqnarray}
	where
	\begin{IEEEeqnarray}{l}
		G^\mathrm{eff}_k(\alpha) 
		:= \max\Big\{C_\mathrm{WT} G(\alpha), \nonumber\\
		C G(\alpha) - \max_{j\in\{0,\ldots,k\}}\left((C-C_\mathrm{WT}) G\left(\tfrac{j}{k}\right) - L\left(\tfrac{j}{k}\right)\right)\Big\}. \IEEEeqnarraynumspace \label{chp:SeqJSC:eqn:GLk}
	\end{IEEEeqnarray}
\end{enumerate}
	Steps $3$ and $4$ are the same as Steps $3$ and $4$ in the proof of \cite[Theorem 2]{HFTM21} and follow the same steps.

\textbf{Proof of Step \ref{chp:SeqJSC:itm:ClasConv}}:
For any $k,n\in\mathbb{N}$, $\ell\in\{0, \ldots, k-1\}$, and $j\in\{\ell, \ldots, k\}$, we have
\begin{IEEEeqnarray}{rCl}
	n k C \sum_{i=j+1}^k{r_i}
	&\stackrel{(a)}{\geq}& \sum_{i=j+1}^k\sum_{m=m_{i-1}+1}^{m_i} \I{U_m}{V_m} \nonumber\\
	&\stackrel{(b)}{\geq}& \Ib{U_{m_j+1}^{m_k}}{V_{m_j+1}^{m_k}} \nonumber\\
	&\stackrel{(c)}{\geq}& \Ib{U_{m_j+1}^{m_k}}{V_{m_j+1}^{m_k} \vert V^{m_j}} \nonumber\\
	&\stackrel{(d)}{\geq}& \Ib{X_{n\ell+1}^{nk}}{V_{m_j+1}^{m_k} \vert V^{m_j}}, \label{chp:SeqJSC:eqn:sumR>I+I}
\end{IEEEeqnarray}
where $(a)$ follows from the definition of $r_i$ (see Definition \ref{chp:SeqJSC:def:Code}) and the capacity of a channel in \eqref{eqn:ChannelC};
$(b)$ follows from the fact that the channel $p(v|u)$ is a DMC;
$(c)$ follows from the Markov chain $V^{m_j} \to U_{m_j+1}^{m_k} \to V_{m_j+1}^{m_k}$;
and $(d)$ follows from the Markov chain $\big(X_{n\ell+1}^{nk}, V^{m_j}\big) \to U_{m_j+1}^{m_k} \to V_{m_j+1}^{m_k}$ and the data processing inequality (see Fig. \ref{fig:system}).
Note that \eqref{chp:SeqJSC:eqn:sumR>I+I} is also valid for $\ell=0$ because $m_0 = 0$.
We define, for $i\in\{1,\ldots,k\}$,
\begin{equation} \label{chp:SeqJSC:eqn:Rtn}
	\begin{cases}
		\tilde{R}_i^{(n)}
		:= \frac{1}{k} R\big(\Eb{d(X_{n (i-1)+1}^{n i}, \hat{X}_{n (i-1)+1}^{n i})}\big), \\
		\tilde{R}_i := \liminf_{n\to\infty}{\tilde{R}_i^{(n)}},
	\end{cases}
\end{equation}
which $R(\cdot)$ is the rate-distortion function for $X \sim p(x)$ and distortion function $d$ (see Definition \ref{def:classical R(D)}).
Later, we show that
\begin{IEEEeqnarray}{l}
	\Ib{X^{nk}}{V_{m_j+1}^{m_k} \vert V^{m_j}}
	\geq n k \sum_{i=j+1}^k{\tilde{R}_i^{(n)}}, \nonumber\\
	\qquad j=0,\ldots,k-1, \label{chp:SeqJSC:eqn:I>sumRt} \\
	\Ib{X_{n\ell+1}^{nk}}{V_{m_j+1}^{m_k} \vert V^{m_j}} \nonumber\\ 
	\qquad\geq nk\sum_{i=\ell}^j{\tilde{R}_i^{(n)}} - nk L\left(\tfrac{j}{k}\right) - n k C_\mathrm{WT} \sum_{i=\ell}^j{r_j}, \nonumber\\
	\qquad\qquad \ell=0,\ldots,k-1,
	\quad j=\ell,\ldots,k. \label{chp:SeqJSC:eqn:I<max0,R-L}
\end{IEEEeqnarray}

\emph{Proof of \eqref{chp:SeqJSC:eqn:G-Gj/k>sumRhat}}:
It follows from \eqref{chp:SeqJSC:eqn:sumR>I+I}, \eqref{chp:SeqJSC:eqn:I>sumRt} utilizing the fact that the left-hand side of \eqref{chp:SeqJSC:eqn:I>sumRt} is the right-hand side of \eqref{chp:SeqJSC:eqn:sumR>I+I} for $\ell=0$, and by taking $\liminf$ as $n\to\infty$.

\emph{Proof of \eqref{chp:SeqJSC:eqn:L>sumRhat}}:
It follows from \eqref{chp:SeqJSC:eqn:sumR>I+I}, \eqref{chp:SeqJSC:eqn:I<max0,R-L}, and taking $\liminf$ as $n\to\infty$.

\emph{Proof of \eqref{chp:SeqJSC:eqn:avgDhatR<d}}:
From Definition \ref{def:classical R(D)}, we have
\begin{IEEEeqnarray}{rCl} 
	\frac{1}{k} \sum_{i=1}^k {D\big(k \tilde{R}_i^{(n)}\big)}
	&\stackrel{(a)}{\leq}& \frac{1}{k} \sum_{i=1}^k{\Eb{d(X_{n (i-1)+1}^{n i}, \hat{X}_{n (i-1)+1}^{n i})}} \nonumber\\
	&\stackrel{(b)}{=}& \E{d(X^{n k},\hat{X}^{n k})}, \label{chp:SeqJSC:eqn:avgDn<d}
\end{IEEEeqnarray}
where $D(\cdot)$ is the distortion-rate function;
$\tilde{R}_i^{(n)}$ was defined in \eqref{chp:SeqJSC:eqn:Rtn};
$(a)$ follows from the definition of $D(\cdot)$;
and $(b)$ follows from \eqref{chp:SeqJSC:eqn:partiald}.
From Remark \ref{cor:R(D)convex}, $D(\cdot)$ is continuous and non-increasing.
As a result, we obtain
$
	\limsup_{n\to\infty}{D(k \tilde{R}_i^{(n)})}
	= D(k \liminf_{n\to\infty}{\tilde{R}_i^{(n)}})
	= D(k \tilde{R}_i).
$
Hence, by taking $\limsup$ from both sides of \eqref{chp:SeqJSC:eqn:avgDn<d}, we obtain that
\begin{IEEEeqnarray*}{rCl}
	\frac{1}{k} \sum_{i=1}^k{D(k \tilde{R}_i)}
	&\leq& \limsup_{n\to\infty}\E{d(X^{n k},\hat{X}^{n k})} \\
	&\stackrel{(a)}{\leq}& \bar{d},
\end{IEEEeqnarray*}
where $(a)$ follows from the fact that, in the converse part, we assume that $\crdf$ $G$ is achievable (see Definition \ref{chp:SeqJSC:def:Achievable R(D)}).

Thus, Step \ref{chp:SeqJSC:itm:ClasConv} is proved.
Now, it only remains to prove \eqref{chp:SeqJSC:eqn:I>sumRt} and \eqref{chp:SeqJSC:eqn:I<max0,R-L}.

\emph{Proof of \eqref{chp:SeqJSC:eqn:I>sumRt}}:
The claim follows from the following sequence of inequalities:
\begin{IEEEeqnarray}{rCl}
	\Ib{X^{n k}}{V_{m_j+1}^{m_k} \vert V^{m_j}}
	&\geq& \Ib{X_{n j + 1}^{n k}}{V_{m_j+1}^{m_k} \vert V^{m_j}} \nonumber\\
	&\stackrel{(a)}{=}& \Ib{X_{n j + 1}^{n k}}{V^{m_k}} \nonumber\\
	&\stackrel{(b)}{=}& \Ib{X_{n j + 1}^{n k}}{V^{m_k}, \hat{X}_{n j + 1}^{n k}} \nonumber\\
	&\geq& \Ib{X_{n j + 1}^{n k}}{\hat{X}_{n j + 1}^{n k}}, \label{chp:SeqJSC:eqn:convFirstIneq1}
\end{IEEEeqnarray}
where $(a)$ follows since $X^{nj}$ is independent of $X_{nj+1}^{n k}$, as a result, according to Definition \ref{chp:SeqJSC:def:Code}, $V^{m_j}$ is independent of $X_{nj+1}^{n k}$ (see Fig. \ref{fig:system});
and $(b)$ follows from the Markov chain $X_{nj+1}^{nk} \to V^{m_k} \to \hat{X}_{nj+1}^{n k}$.
Later, we show that
\begin{IEEEeqnarray}{l}
	\Ib{X_{n j_1 + 1}^{n j_2}}{\hat{X}_{n j_1 + 1}^{n j_2}}
	\geq \sum_{i=j_1+1}^{j_2}{n k \tilde{R}_i^{(n)}}, \nonumber\\
	\qquad 0\leq j_1 \leq j_2 \leq k. \label{chp:SeqJSC:eqn:I(Xn;hatXn)=sumI(xi,hatXi)}
\end{IEEEeqnarray}
Hence, \eqref{chp:SeqJSC:eqn:I>sumRt} follows from \eqref{chp:SeqJSC:eqn:convFirstIneq1} and \eqref{chp:SeqJSC:eqn:I(Xn;hatXn)=sumI(xi,hatXi)}, by selecting $j_1 = j$ and $j_2=k$.
Now, we prove \eqref{chp:SeqJSC:eqn:I(Xn;hatXn)=sumI(xi,hatXi)}.

\emph{Proof of \eqref{chp:SeqJSC:eqn:I(Xn;hatXn)=sumI(xi,hatXi)}}:
\begin{IEEEeqnarray*}{rCl}
	\IEEEeqnarraymulticol{3}{l}{\Ib{X_{n j_1 + 1}^{n j_2}}{\hat{X}_{n j_1 + 1}^{n j_2}}} \\
	\qquad&\stackrel{(a)}{=}& \sum_{i=nj_1+1}^{n j_2}{\H{X_i} - \Hb{X_i \big\vert X_{n j_1+1}^{i-1}, \hat{X}_{nj_1+1}^{n j_2}}} \\
	&\geq& \sum_{i=nj_1+1}^{n j_2}{\Ib{X_i}{\hat{X}_i}} \\
	&\stackrel{(b)}{\geq}& \sum_{i=nj_1+1}^{n j_2}{R\big(\Eb{d(X_i, \hat{X}_i)}\big)} \\
	&=& \sum_{i=j_1+1}^{j_2}
	{\sum_{i'=n (i-1)+1}^{n i}{R\big(\Eb{d(X_{i'}, \hat{X}_{i'})}\big)}} \\
	&\stackrel{(c)}{\geq}& \sum_{i=j_1+1}^{j_2}
	{n R\big(\Eb{d(X_{n (i-1)+1}^{n i}, \hat{X}_{n (i-1)+1}^{n i})}\big)} \\
	&=& \sum_{i=j_1+1}^{j_2}{n k \tilde{R}_i^{(n)}},
\end{IEEEeqnarray*}
where $(a)$ follows since $X^n$ is $\iid$;
$(b)$ follows based on the definition of $R(\cdot)$ in \cite[Theorem 3.5]{ElGamal11};
and $(c)$ follows from the convexity of $R(D)$ (see Remark \ref{cor:R(D)convex}).

\emph{Proof of \eqref{chp:SeqJSC:eqn:I<max0,R-L}}:
The claim follows from the following sequence of inequalities:
\begin{IEEEeqnarray}{rCl}
	\IEEEeqnarraymulticol{3}{l}{\Ib{X_{n\ell+1}^{nk}}{V_{m_j+1}^{m_k} \vert V^{m_j}}} \nonumber\\
	\qquad&=& \Ib{X_{n\ell+1}^{nk}}{V^{m_k}}
	- \Ib{X_{n\ell+1}^{nk}}{V^{m_j}} \nonumber\\
	&\stackrel{(a)}{=}& \Ib{X_{n\ell+1}^{nk}}{V^{m_k}}
	- \Ib{X_{n\ell+1}^{nj}}{V_{m_\ell+1}^{m_j}} \nonumber\\
	&\stackrel{(b)}{=}& \Ib{X_{n\ell+1}^{nk}}{V^{m_k}, \hat{X}_{n\ell+1}^{nk}}
	- \Ib{X_{n\ell+1}^{nj}}{V_{m_\ell+1}^{m_j}} \nonumber\\
	&\geq& \Ib{X_{n\ell+1}^{nk}}{\hat{X}_{n\ell+1}^{nk}}
	- \Ib{X_{n\ell+1}^{nj}}{V_{m_\ell+1}^{m_j}}, \label{chp:SeqJSC:eqn:I(Xj;V-j|Vj,X-j)>I(Xj;hXj)-I(Xj;Vj)}
\end{IEEEeqnarray}
where, $(a)$ follows since $X^{nj}$ is independent of $X_{nj+1}^{n k}$, as a result, according to Definition \ref{chp:SeqJSC:def:Code}, $(X_{n\ell+1}^{nj}, V_{m_\ell+1}^{m_j})$, $V^{m_\ell}$, and $X_{nj+1}^{n k}$ are mutually independent;
and $(b)$ follows from the Markov chain $X_{n\ell+1}^{nj} \to V^{m_k} \to \hat{X}_{n\ell+1}^{nj}$ (see Definition \ref{chp:SeqJSC:def:Code}).
Later we show that
\begin{equation} \label{chp:SeqJSC:eqn:WTConvTrick}
	\Ib{X_{n\ell+1}^{nj}}{V_{m_\ell+1}^{m_j}} 
	\leq n k C_\mathrm{WT} \sum_{i=\ell+1}^j{r_j} + nk L\left(\tfrac{j}{k}\right).
\end{equation}
Hence, from \eqref{chp:SeqJSC:eqn:I(Xj;V-j|Vj,X-j)>I(Xj;hXj)-I(Xj;Vj)} and \eqref{chp:SeqJSC:eqn:WTConvTrick}, we obtain
\begin{IEEEeqnarray*}{rCl}
	\IEEEeqnarraymulticol{3}{l}{\Ib{X_{n\ell+1}^{nk}}{V_{m_j+1}^{m_k} \vert V^{m_j}}} \\
	\qquad&\geq& - n k L\left(\tfrac{j}{k}\right) 
	- n k C_\mathrm{WT} \sum_{i=\ell+1}^j{r_j}
	+ \Ib{X_{n\ell+1}^{nk}}{\hat{X}_{n\ell+1}^{nk}} \\
	&\stackrel{(a)}{\geq}& - n k L\left(\tfrac{j}{k}\right)
	- n k C_\mathrm{WT} \sum_{i=\ell+1}^j{r_j} 
	+ n k \sum_{i=\ell+1}^k{\tilde{R}_i^{(n)}}, \label{chp:SeqJSC:eqn:I<R(E[d])}
\end{IEEEeqnarray*}
where $(a)$ follows from \eqref{chp:SeqJSC:eqn:I(Xn;hatXn)=sumI(xi,hatXi)}, for $j_1=\ell$ and $j_2=k$.
Thus, it only remains to prove \eqref{chp:SeqJSC:eqn:WTConvTrick}.

\emph{Proof of \eqref{chp:SeqJSC:eqn:WTConvTrick}}:
The claim follows from the following sequence of inequalities:
\begin{IEEEeqnarray*}{rCl}
	\IEEEeqnarraymulticol{3}{l}{\Ib{X_{n\ell+1}^{nj}}{V_{m_\ell+1}^{m_j}}} \\
	\quad&\stackrel{(a)}{\leq}& \Ib{X_{n\ell+1}^{nj}}{V_{m_\ell+1}^{m_j}} - \Ib{X^{nj}}{Z^{m_j}} +nk L\left(\tfrac{j}{k}\right) \\
	&\leq& \Ib{X_{n\ell+1}^{nj}}{V_{m_\ell+1}^{m_j}} - \Ib{X_{n\ell+1}^{nj}}{Z_{m_\ell+1}^{m_j}} +nk L\left(\tfrac{j}{k}\right) \\
	&=& \sum_{i=m_\ell+1}^{m_j} \Ib{X_{n\ell+1}^{nj}}{V_i \mid V_{m_\ell+1}^{i-1}} \\
	&&- \Ib{X_{n\ell+1}^{nj}}{Z_i \mid Z_{i+1}^{m_j}} \\
	&&+ nk L\left(\tfrac{j}{k}\right) \\
	&\stackrel{(b)}{=}& \sum_{i=m_\ell+1}^{m_j} \Ib{X_{n\ell+1}^{nj}, Z_{i+1}^{m_j}}{V_i \mid V_{m_\ell+1}^{i-1}} \\
	&&\qquad\quad- \Ib{X_{n\ell+1}^{nj}, V_{m_\ell+1}^{i-1}}{Z_i \mid Z_{i+1}^{m_j}} \\
	&&+ nk L\left(\tfrac{j}{k}\right) \\
	&\stackrel{(c)}{=}& \sum_{i=m_\ell+1}^{m_j} 
	\Ib{X_{n\ell+1}^{nj}}{V_i \mid V_{m_\ell+1}^{i-1}, Z_{i+1}^{m_j}} \\
	&&\qquad\quad- \Ib{X_{n\ell+1}^{nj}}{Z_i \mid V_{m_\ell+1}^{i-1}, Z_{i+1}^{m_j}} \\
	&&+ nk L\left(\tfrac{j}{k}\right) \\
	&\stackrel{(d)}{=}& \sum_{i=m_\ell+1}^{m_j} \Ib{W_i}{V_i \mid W'_i} - \Ib{W_i}{Z_i \mid W'_i} \\
	&&+ nk L\left(\tfrac{j}{k}\right) \\
	&\stackrel{(e)}{=}& (m_j-m_\ell) \big[\Ib{W_Q}{V_Q \mid W'_Q, Q} \\
	&& - \Ib{W_Q}{Z_Q \mid W'_Q, Q} \big]
	+ nk L\left(\tfrac{j}{k}\right) \\
	&\stackrel{(f)}{=}& (m_j-m_\ell) \big[\Ib{W}{V \mid W'} \\
	&& - \Ib{W}{Z \mid W'} \big] + nk L\left(\tfrac{j}{k}\right) \\
	&\leq& (m_j-m_\ell) \\
	&&\times\max_{w'}\left[\Ib{W}{V \mid W'=w'} - \Ib{W}{Z \mid W'=w'} \right] \\
	&&+ nk L\left(\tfrac{j}{k}\right) \\
	&\stackrel{(g)}{\leq}& (m_j-m_\ell) C_\mathrm{WT} + nk L\left(\tfrac{j}{k}\right),
\end{IEEEeqnarray*}
where $(a)$ follows from \eqref{chp:SeqJSC:eqn:mainI<L};
$(b)$ and $(c)$ follow from Csisz\'ar sum identity \cite[p. 25]{ElGamal11};
$(d)$ follows by defining $W'_i := \big(V_{m_\ell+1}^{i-1}, Z_{i+1}^{m_j}\big)$ and $W_i := \big(X_{n\ell+1}^{nj}, V_{m_\ell+1}^{i-1}, Z_{i+1}^{m_j}\big)$;
$(e)$ follows by defining $Q\sim\mathrm{Unif}\{m_{\ell+1},\ldots,m_j\}$ independent of $(X^{nk}, U^{m_k}, V^{m_k}, Z^{m_k}, \hat{X}^{nk})$;
$(f)$ follows by defining $W:=(W_Q,Q)$ and $W' := (W'_Q,Q)$; as a result, we have the Markov chain $W' \to W \to U \to (V, Z)$ because $W'$ is a function of $W$;
and $(g)$ follows from the definition of the capacity of wiretap channel (see Theorem \ref{thm:bg:wt:sc}).
Hence, \eqref{chp:SeqJSC:eqn:WTConvTrick} follows from the definition of $r_i$ and $m_i$ (see Definition \ref{chp:SeqJSC:def:Code}).

\textbf{Proof of Step \ref{chp:SeqJSC:itm:decR}}:
We define $\{S_i\}_{i=1}^k$ as the sorted permutation of $\{\tilde{R}_i\}_{i=1}^k$ in descending order.
Hence, we obtain
\begin{IEEEeqnarray}{l}
	S_1 \geq \cdots \geq S_k, \label{chp:SeqJSC:eqn:yr1>yrn} \\
	\sum_{i=1}^j {S_i} 
	\geq \sum_{i=1}^j {\tilde{R}_i}
	\qquad j=1,\ldots,k-1, \label{chp:SeqJSC:eqn:sumyi<sumRt} \\
	\sum_{i=1}^k {S_i} 
	= \sum_{i=1}^k {\tilde{R}_i}, \label{chp:SeqJSC:eqn:sumSi=sumRt} \\
	\sum_{i=1}^k {D(k S_i)} 
	= \sum_{i=1}^k {D(k \tilde{R}_i)}. \label{chp:SeqJSC:eqn:avgDhatyr<d}
\end{IEEEeqnarray}
Next, we define the sequence $\tilde{R}'_j$, for $j=1,\ldots,k$, as
\begin{equation} \label{chp:SeqJSC:eqn:def:tildeR'i}
	\tilde{R}'_j =
	\begin{cases}
		S_1 + G^\mathrm{eff}_k(1) - \sum_{i=1}^k S_i & j = 1, \\
		S_j & j=2, \ldots, k.
	\end{cases}
\end{equation}
Later, we show that
\begin{IEEEeqnarray}{l}
	G_k^\mathrm{eff}(1)  
	= C G(\alpha) - c, \label{chp:SeqJSC:eqn:alphabarGeff} \\
	\sum_{i=1}^k{\tilde{R}_i}
	\leq G^\mathrm{eff}_k(1), \label{chp:SeqJSC:eqn:ctilde>cnormal}
\end{IEEEeqnarray}
where
\begin{equation} \label{chp:SeqJSC:eqn:Constc}
	c := 
	\max_{j\in\{0,\ldots,k\}}(C-C_\mathrm{WT}) G\left(\tfrac{j}{k}\right) - L\left(\tfrac{j}{k}\right).
\end{equation}
Note that, from \eqref{chp:SeqJSC:eqn:sumSi=sumRt}, \eqref{chp:SeqJSC:eqn:def:tildeR'i}, and \eqref{chp:SeqJSC:eqn:ctilde>cnormal}, we have $\tilde{R}'_1 \geq S_1 \geq 0$;
as a result, $\tilde{R}'_1$ is a valid rate.

\emph{Proof of \eqref{chp:SeqJSC:eqn:R1>Rn}}:
From \eqref{chp:SeqJSC:eqn:yr1>yrn} and \eqref{chp:SeqJSC:eqn:def:tildeR'i}, we obtain that $S_1\geq\tilde{R}'_2\geq\ldots\geq\tilde{R}'_k$.
Thus, \eqref{chp:SeqJSC:eqn:R1>Rn} follows from the fact that $\tilde{R}'_1 \geq S_1$.

\emph{Proof of \eqref{chp:SeqJSC:eqn:sumRt''>GL}}:
We can write, for $j=1,\ldots,k-1$,
\begin{IEEEeqnarray}{rCl} 
	\sum_{i=1}^j \tilde{R}'_i
	&\stackrel{(a)}{=}& G_k^\mathrm{eff}(1) - \sum_{i=1}^k S_i
	+ \sum_{i=1}^j S_i \nonumber\\
	&=& G_k^\mathrm{eff}(1) - \sum_{i=j+1}^k S_i \nonumber\\
	&\stackrel{(b)}{\geq}& G_k^\mathrm{eff}(1) - \sum_{i=j+1}^{k} \tilde{R}_i, \label{chp:SeqJSC:eqn:sumRt'>GL1-sumRt_}
\end{IEEEeqnarray}
where $(a)$ follows from \eqref{chp:SeqJSC:eqn:def:tildeR'i}
and $(b)$ follows from \eqref{chp:SeqJSC:eqn:sumyi<sumRt} and \eqref{chp:SeqJSC:eqn:sumSi=sumRt}.
Therefore, it is sufficient to show that
\begin{equation} \label{chp:SeqJSC:eqn:conv step2, main1}
	\sum_{i=j+1}^{k} \tilde{R}_i
	\leq G_k^\mathrm{eff}(1) - G_k^\mathrm{eff}\left(\tfrac{j}{k}\right).
\end{equation}
To this end, we consider the following cases:
\begin{itemize}
	\item From \eqref{chp:SeqJSC:eqn:G-Gj/k>sumRhat}, we obtain,
		\begin{IEEEeqnarray}{rCl}
			\sum_{i=j+1}^{k} \tilde{R}_i
			&\leq& C \sum_{i=j+1}^k r_i \nonumber\\
			&\stackrel{(a)}{=}& C\left(G(1) - G\left(\tfrac{j}{k}\right)\right) \nonumber\\
			&=& C G(1) - c - \left(C G\left(\tfrac{j}{k}\right) - c\right) \nonumber\\
			&\stackrel{(b)}{=}& G_k^\mathrm{eff}(1) - \left(C G\left(\tfrac{j}{k}\right) - c\right), \label{chp:SeqJSC:eqn:GL1-GLj>sumRt_}
		\end{IEEEeqnarray}
		where $(a)$ follows from Definition \ref{chp:SeqJSC:def:Code};
		and $(b)$ follows from \eqref{chp:SeqJSC:eqn:alphabarGeff}.

	\item From \eqref{chp:SeqJSC:eqn:L>sumRhat}, we have for $j'\geq j$ and $j=0,\ldots,k-1$,
		\begin{IEEEeqnarray*}{rCl}
			\sum_{i=j+1}^k{\tilde{R}_i}
			&\leq& L\left(\tfrac{j'}{k}\right) 
			+ C_\mathrm{WT}\sum_{i=j+1}^{j'}{r_i}
			+ C \sum_{i=j'+1}^k {r_i} \\
			&=& L\left(\tfrac{j'}{k}\right) 
			- (C-C_\mathrm{WT}) \sum_{i=1}^{j'}{r_i} \\
			&&+ (C-C_\mathrm{WT}) \sum_{i=1}^{j'}{r_i} \\
			&&+ C_\mathrm{WT}\sum_{i=j+1}^{j'}{r_i}
			+ C \sum_{i=j'+1}^k {r_i} \\
			&=& L\left(\tfrac{j'}{k}\right) 
			- (C-C_\mathrm{WT}) \sum_{i=1}^{j'}{r_i} \\
			&&- C_\mathrm{WT}\sum_{i=1}^{j}{r_i}
			+ C \sum_{i=1}^k {r_i}.
		\end{IEEEeqnarray*}
		Hence, by minimizing the right-hand side over $j'$, we have
		\begin{IEEEeqnarray}{rCl}
			\sum_{i=j+1}^k{\tilde{R}_i}
			&\leq& C \sum_{i=1}^k {r_i}
			- C_\mathrm{WT}\sum_{i=1}^{j}{r_i} \nonumber\\
			&&- \max_{j' \geq j} \left((C-C_\mathrm{WT}) \sum_{i=1}^{j'}{r_i}
			- L\left(\tfrac{j'}{k}\right)\right) \nonumber\\
			&\leq& C \sum_{i=1}^k {r_i}
			- C_\mathrm{WT}\sum_{i=1}^{j}{r_i} \nonumber\\
			&&- \max_{j' \geq 0} \left((C-C_\mathrm{WT}) \sum_{i=1}^{j'}{r_i}
			- L\left(\tfrac{j'}{k}\right)\right) \nonumber\\
			&=& CG(1) - c - C_\mathrm{WT}G\left(\tfrac{j}{k}\right) \nonumber\\
			&\stackrel{(a)}{=}& G_k^\mathrm{eff}(1) - C_\mathrm{WT}G\left(\tfrac{j}{k}\right), \label{chp:SeqJSC:eqn:Step2:ineq3}
		\end{IEEEeqnarray}
		where $c$ was defined in \eqref{chp:SeqJSC:eqn:Constc};
		and $(a)$ follows from \eqref{chp:SeqJSC:eqn:alphabarGeff}.
\end{itemize}
Thus, \eqref{chp:SeqJSC:eqn:conv step2, main1} follows from \eqref{chp:SeqJSC:eqn:GL1-GLj>sumRt_} and \eqref{chp:SeqJSC:eqn:Step2:ineq3}.
Now, it only remains to prove \eqref{chp:SeqJSC:eqn:alphabarGeff} and \eqref{chp:SeqJSC:eqn:ctilde>cnormal}.

\emph{Proof of \eqref{chp:SeqJSC:eqn:alphabarGeff}}:
We have
\begin{IEEEeqnarray*}{rCl}
	\IEEEeqnarraymulticol{3}{l}{C G(1) - c - C_\mathrm{WT} G(1)} \\
	\qquad&\stackrel{(a)}{=}& (C-C_\mathrm{WT}) G(1) 
	- \max_{j\in\{0,\ldots,k\}}(C-C_\mathrm{WT}) G\left(\tfrac{j}{k}\right) \\
	&& - L\left(\tfrac{j}{k}\right) \\
	&\geq& (C-C_\mathrm{WT}) G(1) 
	- \max_{j\in\{0,\ldots,k\}}(C-C_\mathrm{WT}) G\left(\tfrac{j}{k}\right) \\
	&\stackrel{(b)}{=}& 0,
\end{IEEEeqnarray*}
where $(a)$ follows from the definition of $c$ in \eqref{chp:SeqJSC:eqn:Constc};
and $(b)$ follows from the fact that $G(\alpha)$ is a non-decreasing function, so, the maximum occures at $\alpha=1$. 
Thus, \eqref{chp:SeqJSC:eqn:alphabarGeff} follows from \eqref{chp:SeqJSC:eqn:GLk}.

\emph{Proof of \eqref{chp:SeqJSC:eqn:ctilde>cnormal}}:
From \eqref{chp:SeqJSC:eqn:L>sumRhat}, by taking $\ell=0$, we have, for $j=0,\ldots,k$,
\begin{IEEEeqnarray*}{rCl}
	\sum_{i=1}^k{\tilde{R}_i}
	&\leq& L\left(\tfrac{j}{k}\right) 
	+ C_\mathrm{WT}\sum_{i=1}^j{r_i}
	+ C \sum_{i=j+1}^k {r_i}, \\
	&=& C \sum_{i=1}^k {r_i}
	+ L\left(\tfrac{j}{k}\right)
	- (C - C_\mathrm{WT}) \sum_{i=1}^j{r_i}.
\end{IEEEeqnarray*}
By taking the minimum of the right-hand side over $j$, we obtain
\begin{IEEEeqnarray*}{rCl}
	\sum_{i=1}^k{\tilde{R}_i}
	&\leq& C \sum_{i=1}^k {r_i}
	- \max_{j=0,\ldots,k} (C - C_\mathrm{WT}) \sum_{i=1}^j{r_i}
	- L\left(\tfrac{j}{k}\right) \\
	&=& C \sum_{i=1}^k {r_i} - c,
\end{IEEEeqnarray*}
where $c$ was defined in \eqref{chp:SeqJSC:eqn:Constc}.
Thus, \eqref{chp:SeqJSC:eqn:ctilde>cnormal} follows from \eqref{chp:SeqJSC:eqn:alphabarGeff} the definition of $G(\alpha)$ (see Definition \ref{chp:SeqJSC:def:Code}).

\emph{Proof of \eqref{chp:SeqJSC:eqn:totRt''=GL1-GL0}}:
It follows from \eqref{chp:SeqJSC:eqn:def:tildeR'i}.

\emph{Proof of \eqref{chp:SeqJSC:eqn:avgDhatR''<d}}:
Since the distortion-rate function $D(\cdot)$ is non-increasing (see Remark \ref{cor:R(D)convex}), $D(k\tilde{R}'_1) \leq D(k S_1)$ because $\tilde{R}'_1 \geq S_1$.
Besides, for $i=2,\ldots,k$, $D(k \tilde{R}'_i) = D(k S_i)$; as a result, \eqref{chp:SeqJSC:eqn:avgDhatR''<d} follows from \eqref{chp:SeqJSC:eqn:avgDhatyr<d}.

Thus, Step \ref{chp:SeqJSC:itm:decR} is proved and the derivation of the proof is complete.
\qed

\subsection{Proof of Theorem \ref{CHP:SEQJSC:THM:R(D):ACH}} \label{chp:SeqJSC:subsec:prf:R(D):ach}
To prove the achievability, we separate source and channel coding:

\emph{Source Coding}:
From the classical rate-distortion theorem \cite[Theorem 3.5]{ElGamal11}, we have that, for any $k$, there exists a set of encoders $\bar{\varphi}_i^\mathrm{s}$ and decoders $\bar{\psi}_i^\mathrm{s}$, for $i\in\{1,\ldots,k\}$, such that
\begin{IEEEeqnarray}{rCl}
	\bar{\varphi}_i^\mathrm{s} &\colon& \mathcal{X}^n \to \{1,\ldots,2^{\floor{nk \hat{R}_i}}\} \nonumber\\
	&& x_{(i-1)n+1}^{i n} \mapsto \bar{w}_i, \nonumber\\
	\bar{\psi}_i^\mathrm{s} &\colon& \{1,\ldots,2^{\floor{nk \hat{R}_i}}\} \to \hat{\mathcal{X}}^n \nonumber\\
	&& \bar{w}_i \mapsto \hat{x}_{(i-1)n+1}^{i n}, \nonumber\\
	\IEEEeqnarraymulticol{3}{l}{\limsup_{n\to\infty} \E{d(X_{(i-1)n+1}^{i n}, \hat{X}_{(i-1)n+1}^{i n})} \leq D(k \hat{R}_i)}, \label{chp:SeqJSC:inequality2}
\end{IEEEeqnarray}
where $D(\cdot)$ is the distortion-rate function (see Definition \ref{def:classical R(D)}).
We choose
\begin{equation*}
	\hat{R}_i 
	= \hat{G}_\mathrm{in}^\mathrm{eff}\left(\tfrac{i}{k}\right) - \hat{G}_\mathrm{in}^\mathrm{eff}\left(\tfrac{i-1}{k}\right)
	= \int_{\frac{i-1}{k}}^{\frac{i}{k}}{\frac{\ud \hat{G}_\mathrm{in}^\mathrm{eff}}{\ud \alpha}(\alpha) \ud \alpha}.
\end{equation*}
From Remark \ref{cor:R(D)convex}, $D(\cdot)$ is convex.
So, using Jensen's inequality \cite[Theorem 2.6.2]{Cover06}, we have
\begin{equation} \label{chp:SeqJSC:inequality3}
	D\left(k \int_{\frac{i-1}{k}}^{\frac{i}{k}}{\frac{\ud \hat{G}_\mathrm{in}^\mathrm{eff}}{\ud \alpha}(\alpha) \ud \alpha} \right)
	\leq k \int_{\frac{i-1}{k}}^{\frac{i}{k}}{D\left( \frac{\ud \hat{G}_\mathrm{in}^\mathrm{eff}}{\ud \alpha}(\alpha) \right) \ud \alpha}.
\end{equation}
Therefore, from \eqref{chp:SeqJSC:inequality2} and \eqref{chp:SeqJSC:inequality3}, when $n\to\infty$, we have
\begin{IEEEeqnarray*}{rCl}
	\IEEEeqnarraymulticol{3}{l}{\limsup_{n\to\infty}\E{d(X^{k n}, \hat{X}^{k n})}} \\
	\qquad&=& \frac{1}{k} \sum_{i=1}^k{\limsup_{n\to\infty}\Eb{d\big(X_{(i-1)n+1}^{i n}, \hat{X}_{(i-1)n+1}^{i n} \big)}} \\
	&\leq& \frac{1}{k} \sum_{i=1}^k{k \int_{\frac{i-1}{k}}^{\frac{i}{k}}{D\left( \frac{\ud \hat{G}_\mathrm{in}^\mathrm{eff}}{\ud \alpha}(\alpha) \right) \ud \alpha}} \\
	&=& \int_0^1{D\left( \frac{\ud \hat{G}_\mathrm{in}^\mathrm{eff}}{\ud \alpha}(\alpha) \right) \ud\alpha} \\
	&\stackrel{(a)}{\leq}& \bar{d},
\end{IEEEeqnarray*}
where $(a)$ follows from \eqref{chp:SeqJSC:eqn:intDG<d}.
Hence, the distortion constraint \eqref{chp:SeqJSC:eqn:limsupEd<bard} is satisfied if, for each block $i$, the rate $R_i$ is communicated to the receiver.

\emph{Reshaping the Rate Profile}:
From Lemma \ref{chp:SeqJSC:lmm:CRDFMaj}, and the source coding part, we obtain that the rate profile $G^\mathrm{eff}$ is also achievable with given distortion constraint, i.e., for any $k$, there exists a set of encoders $\varphi_i^\mathrm{s}$, for $i\in\{1,\ldots,k\}$, and a decoder $\psi^\mathrm{s}$ such that
\begin{IEEEeqnarray}{rCl}
	\varphi_i^\mathrm{s} &\colon& \mathcal{X}^{i n} \to \left\{1,\ldots,2^{\floor{nk R_i}}\right\} \nonumber\\
	&& x^{i n} \mapsto w_i, \nonumber\\
	\psi^\mathrm{s} &\colon& \left\{1,\ldots,2^{\floor{nk G_\mathrm{in}^\mathrm{eff}(1)}}\right\} \to \hat{\mathcal{X}}^n \label{chp:SeqJSC:eqn:ach:dec}\\
	&& w^k \mapsto \hat{x}^{k n}, \nonumber\\
	\IEEEeqnarraymulticol{3}{l}{\limsup_{n\to\infty} \E{d(X^{k n}, \hat{X}^{k n})} \leq \bar{d},} \label{chp:SeqJSC:eqn:ach:R(D)2}
\end{IEEEeqnarray}
where the fact that $\hat{G}_\mathrm{in}^\mathrm{eff}(1) = G_\mathrm{in}^\mathrm{eff}(1)$, in \eqref{chp:SeqJSC:eqn:ach:dec}, follows from \cite[Lemma 3]{HFTM21},
and
\begin{equation} \label{chp:SeqJSC:eqn:ach:Ri}
	R_i 
	= G_\mathrm{in}^\mathrm{eff}\left(\tfrac{i}{k}\right) - G_\mathrm{in}^\mathrm{eff}\left(\tfrac{i-1}{k}\right).
\end{equation}

\emph{Channel Coding}:
To complete the proof, it suffices to show that, for Block $i$ and $r_i$ given in \eqref{chp:SeqJSC:eqn:def R}, there is an encoding scheme to transmit $W_i$, losslessly, through the corresponding channel $p(z,v|u)$ such that the leakage constraint \eqref{chp:SeqJSC:eqn:mainI<L} is satisfied.
The rate of encoding (bits per channel usage) for Block $i$ is
\begin{equation} \label{chp:SeqJSC:eqn:ach:barR}
	\bar{R}_i 
	:= \lim_{n\to\infty}\frac{\floor{n k R_i}}{\floor{n k r_i}}
	= \frac{R_i}{r_i},
\end{equation}
where $R_i$ and $r_i$ are defined in \eqref{chp:SeqJSC:eqn:ach:Ri} and \eqref{chp:SeqJSC:eqn:def R}, respectively.
Later, we show that
\begin{equation} \label{chp:SeqJSC:eqn:ach:barR<C}
	\bar{R}_i \leq C_2.
\end{equation}
Therefore, we can use the coding scheme in Lemma \ref{chp:SeqJSC:lmm:DWTC} for each block.
Hence, $W_i$ is transmitted, losslessly, and we also have
\begin{IEEEeqnarray}{l}
	\limsup_{n\to\infty}\frac{1}{\floor{n k r_i}}\Ib{W_i}{Z_{m_{i-1}+1}^{m_i}} 
	\leq \max\{0, \ell(\bar{R}_i - C_1)\}. \IEEEeqnarraynumspace \label{chp:SeqJSC:eqn:ach:leakcont1}
\end{IEEEeqnarray}
In order to show \eqref{chp:SeqJSC:eqn:mainI<L}, we can write
\begin{IEEEeqnarray}{rCl}
	\Ib{X^{i n}}{Z^{m_i}}
	&\stackrel{(a)}{\leq}& \Ib{W^i}{Z^{m_i}} \nonumber\\
	&\stackrel{(b)}{=}& \H{Z^{m_i}} - \sum_{j=1}^i{\Hb{Z_{m_{j-1}+1}^{m_j} \mid W_j}} \nonumber\\
	&\leq& \sum_{j=1}^i \Ib{W_j}{Z_{m_{j-1}+1}^{m_j}}, \label{chp:SeqJSC:eqn:ach:blockSep}
\end{IEEEeqnarray}
where $(a)$ follows from the Markov chain $X^{in} \to W^i \to U^{m_i} \to Z^{m_i}$; and
$(b)$ follows from the Markov chain $W^{i-1} \to W_i \to Z_{m_{i-1}+1}^{m_i}$.
Therefore,
\begin{IEEEeqnarray*}{rCl}
	\IEEEeqnarraymulticol{3}{l}{\limsup_{n\to\infty}\frac{1}{n k}\Ib{X^{i n}}{Z^{m_i}}} \\
	\qquad&\stackrel{(a)}{\leq}& \sum_{j=1}^i \limsup_{n\to\infty}\frac{1}{n k}\Ib{W_j}{Z_{m_{j-1}+1}^{m_j}} \\
	&\stackrel{(b)}{\leq}& \sum_{j=1}^i \max\left\{0, \ell r_j (\bar{R}_j - C_1)\right\} \\
	&\stackrel{(c)}{=}& \ell\sum_{j=1}^i \max\left\{0, R_j - r_j C_1\right\} \\
	&\stackrel{(d)}{=}& \ell\sum_{j=1}^i R_j - r_j C_1 \\
	&=& \ell\left(G_\mathrm{in}^\mathrm{eff}\left(\tfrac{i}{k}\right) - C_1 G\left(\tfrac{i}{k}\right)\right) \\
	&=& \ell\max\Big\{0, (C_2-C_1)G\left(\tfrac{i}{k}\right) \\
	&&\qquad\qquad- \sup_{\beta\in[0,1]}(C_2-C_1) G(\beta) - \frac{L(\beta)}{\ell} \Big\} \\
	&\stackrel{(e)}{\leq}& \ell\max\Big\{0, (C_2-C_1)G\left(\tfrac{i}{k}\right) \\
	&&\qquad\qquad -(C_2-C_1) G\left(\tfrac{i}{k}\right) + \frac{L\left(\tfrac{i}{k}\right)}{\ell} \Big\} \\
	&=& L\left(\tfrac{i}{k}\right),
\end{IEEEeqnarray*}
where $(a)$ follows from \eqref{chp:SeqJSC:eqn:ach:blockSep};
$(b)$ follows from \eqref{chp:SeqJSC:eqn:ach:leakcont1};
$(c)$ follows from \eqref{chp:SeqJSC:eqn:ach:barR};
$(d)$ follows from the fact that $G^\mathrm{eff}(\alpha)$ can be written as the sum of two non-decreasig functions:
\begin{IEEEeqnarray*}{l}
	C_1 G(\alpha), \\
	\max\Big\{0, \\
	\quad(C_2-C_1) G(\alpha) - \sup_{\beta\in[0,1]}(C_2-C_1) G(\beta) - L(\beta)/\ell\Big\};
\end{IEEEeqnarray*}
 as a result, 
$G_\mathrm{in}^\mathrm{eff}(\alpha_1) - G_\mathrm{in}^\mathrm{eff}(\alpha_2) \geq C_1 G(\alpha_1) - C_1 G(\alpha_2)$ for any $\alpha_1 \geq \alpha_2$ that results in $R_j \geq r_j C_1$;
and $(e)$ follows by choosing $\beta = i/k$.
Thus, the achievability is derived.
It only remains to prove \eqref{chp:SeqJSC:eqn:ach:barR<C}.

\emph{Proof of \eqref{chp:SeqJSC:eqn:ach:barR<C}}:
We need to show, for $ i=1,\ldots,k$.
\begin{equation} \label{chp:SeqJSC:eqn:ach:Cr-r>0}
	C_2 r_i - R_i \geq 0.
\end{equation}
We define
\begin{IEEEeqnarray}{l}
	c' := \sup_{\beta\in[0,1]}\left((C_2-C_1) G(\beta) - \frac{L(\beta)}{\ell}\right), \nonumber\\
	G'(\alpha) := \max\left\{0, (C_2 - C_1) G(\alpha) - c'\right\}. \label{chp:SeqJSC:eqn:ach:extra1}
\end{IEEEeqnarray}
Hence, 
\begin{IEEEeqnarray*}{rCl}
	\IEEEeqnarraymulticol{3}{l}{C_2 r_i - R_i} \\
	\quad&=&  C_2 \left[G\left(\tfrac{i}{k}\right) - G\left(\tfrac{i-1}{k}\right)\right]
	- \left[G_\mathrm{in}^\mathrm{eff}\left(\tfrac{i}{k}\right) - G_\mathrm{in}^\mathrm{eff}\left(\tfrac{i-1}{k}\right)\right] \\
	&\stackrel{(a)}{=}& C_2 \left[G\left(\tfrac{i}{k}\right) - G\left(\tfrac{i-1}{k}\right)\right] \\
	&&- C_1 \left[G\left(\tfrac{i}{k}\right) - G\left(\tfrac{i-1}{k}\right)\right]
	- \left[G'\left(\tfrac{i}{k}\right) - G'\left(\tfrac{i-1}{k}\right)\right] \\
	&=& (C_2 - C_1) \left[G\left(\tfrac{i}{k}\right) - G\left(\tfrac{i-1}{k}\right)\right] \\
	&&- \left[G'\left(\tfrac{i}{k}\right) - G'\left(\tfrac{i-1}{k}\right)\right],
\end{IEEEeqnarray*}
where $(a)$ follows from the fact that $G_\mathrm{in}^\mathrm{eff}(\alpha) = C_1 G(\alpha) + G'(\alpha)$.
If $G'(i/k) = 0$, then $G'((i-1)/k)=0$ because $G'(\alpha)$ is non-decreasing; as a result, \eqref{chp:SeqJSC:eqn:ach:Cr-r>0} follows.
Otherwise, we can write
\begin{IEEEeqnarray*}{rCl}
	C_2 r_i - R_i
	&=& (C_2 - C_1) \left[G\left(\tfrac{i}{k}\right) - G\left(\tfrac{i-1}{k}\right)\right] \\
	&&- \left[(C_2 - C_1) G\left(\tfrac{i}{k}\right) - c - G'\left(\tfrac{i-1}{k}\right)\right] \\
	&\stackrel{(a)}{\geq}& (C_2 - C_1) \left[G\left(\tfrac{i}{k}\right) - G\left(\tfrac{i-1}{k}\right)\right] \\
	&& - \left[(C_2 - C_1) G\left(\tfrac{i}{k}\right) - (C_2 - C_1) G\left(\tfrac{i-1}{k}\right)\right] \\
	&=& 0,
\end{IEEEeqnarray*}
where $(a)$ follows from \eqref{chp:SeqJSC:eqn:ach:extra1}.
\qed

\section{Conclusion} \label{sec:conclusion}
In this work, the results of Ghourchian et al. \cite{HFTM21} were extended to joint source-channel coding with eavesdropping. 
The sequential encoding process was characterized using cumulative rate distribution functions ($\crdf$) and the security constraint was incorporated using cumulative leakage distribution functions ($\cldf$). The information leakage was defined based on the mutual information between the source and the output of the wiretap channel to the eavesdropper. Inner and outer bounds on the achievable $\crdf$ for a given source and $\cldf$ were derived and it was shown that the bounds were tight when the distribution that achieved the capacity of the wiretap channel was the same as the one that achieved the capacity of the channel.

For the future works, the results can be extended to find the whole region and tighter bounds in general.

\bibliographystyle{IEEEtran}
\bibliography{IEEEabrv,myref}

\begin{thebibliography}{10}
\providecommand{\url}[1]{#1}
\csname url@samestyle\endcsname
\providecommand{\newblock}{\relax}
\providecommand{\bibinfo}[2]{#2}
\providecommand{\BIBentrySTDinterwordspacing}{\spaceskip=0pt\relax}
\providecommand{\BIBentryALTinterwordstretchfactor}{4}
\providecommand{\BIBentryALTinterwordspacing}{\spaceskip=\fontdimen2\font plus
\BIBentryALTinterwordstretchfactor\fontdimen3\font minus
  \fontdimen4\font\relax}
\providecommand{\BIBforeignlanguage}[2]{{%
\expandafter\ifx\csname l@#1\endcsname\relax
\typeout{** WARNING: IEEEtran.bst: No hyphenation pattern has been}%
\typeout{** loaded for the language `#1'. Using the pattern for}%
\typeout{** the default language instead.}%
\else
\language=\csname l@#1\endcsname
\fi
#2}}
\providecommand{\BIBdecl}{\relax}
\BIBdecl

\bibitem{HFTM21}
H.~Ghourchian, P.~A. Stavrou, T.~J. Oechtering, and M.~Skoglund, ``Secure block
  source coding with sequential encoding,'' \emph{IEEE Journal on Selected
  Areas in Information Theory}, vol.~2, no.~1, pp. 32--48, 2021.

\bibitem{Shannon59}
C.~E. Shannon, ``Coding theorems for a discrete source with a fidelity
  criterion,'' \emph{IRE Nat. Conv. Rec}, vol.~4, no.~1, pp. 325--350, 1959.

\bibitem{Neuhoff82}
D.~Neuhoff and R.~Gilbert, ``Causal source codes,'' \emph{{IEEE} Trans. Inf.
  Theory}, vol.~28, no.~5, pp. 701--713, 1982.

\bibitem{Weissman05}
T.~{Weissman} and N.~{Merhav}, ``On causal source codes with side
  information,'' \emph{{IEEE} Trans. Inf. Theory}, vol.~51, no.~11, pp.
  4003--4013, 2005.

\bibitem{Linder01}
T.~Linder and G.~Lagosi, ``A zero-delay sequential scheme for lossy coding of
  individual sequences,'' \emph{{IEEE} Trans. Inf. Theory}, vol.~47, no.~6, pp.
  2533--2538, 2001.

\bibitem{Stavrou18}
P.~A. Stavrou, J.~{\O}stergaard, and C.~D. Charalambous, ``Zero-delay rate
  distortion via filtering for vector-valued {Gaussian} sources,'' \emph{IEEE
  J. Sel. Topics Signal Process.}, vol.~12, no.~5, pp. 841--856, 2018.

\bibitem{Tanaka18}
T.~{Tanaka}, P.~M. {Esfahani}, and S.~K. {Mitter}, ``{LQG} control with minimum
  directed information: Semidefinite programming approach,'' \emph{{IEEE}
  Trans. Autom. Control}, vol.~63, no.~1, pp. 37--52, 2018.

\bibitem{Akyol14}
E.~Akyol, K.~B. Viswanatha, K.~Rose, and T.~A. Ramstad, ``On zero-delay
  source-channel coding,'' \emph{{IEEE} Trans. Inf. Theory}, vol.~60, no.~12,
  pp. 7473--7489, 2014.

\bibitem{Matloub06}
S.~Matloub and T.~Weissman, ``Universal zero-delay joint source--channel
  coding,'' \emph{{IEEE} Trans. Inf. Theory}, vol.~52, no.~12, pp. 5240--5250,
  2006.

\bibitem{Merhav03}
N.~Merhav and I.~Kontoyiannis, ``Source coding exponents for zero-delay coding
  with finite memory,'' \emph{{IEEE} Trans. Inf. Theory}, vol.~49, no.~3, pp.
  609--625, 2003.

\bibitem{Viswanathan00}
H.~Viswanathan and T.~Berger, ``Sequential coding of correlated sources,''
  \emph{{IEEE} Trans. Inf. Theory}, vol.~46, no.~1, pp. 236--246, 2000.

\bibitem{Ma11}
N.~Ma and P.~Ishwar, ``On delayed sequential coding of correlated sources,''
  \emph{{IEEE} Trans. Inf. Theory}, vol.~57, no.~6, pp. 3763--3782, 2011.

\bibitem{Shannon49}
C.~E. Shannon, ``Communication theory of secrecy systems,'' \emph{Bell Sys.
  Tech. J.}, vol.~28, no.~4, pp. 656--715, 1949.

\bibitem{Wyner75}
A.~D. {Wyner}, ``The wire-tap channel,'' \emph{Bell Sys. Tech. J.}, vol.~54,
  no.~8, pp. 1355--1387, 1975.

\bibitem{Yamamoto97}
H.~{Yamamoto}, ``Rate-distortion theory for the {Shannon} cipher system,''
  \emph{{IEEE} Trans. Inf. Theory}, vol.~43, no.~3, pp. 827--835, 1997.

\bibitem{Prabhakaran07}
V.~{Prabhakaran} and K.~{Ramchandran}, ``On secure distributed source coding,''
  in \emph{2007 IEEE Information Theory Workshop}, 2007, pp. 442--447.

\bibitem{Gunduz08}
D.~{Gunduz}, E.~{Erkip}, and H.~V. {Poor}, ``Secure lossless compression with
  side information,'' in \emph{2008 IEEE Information Theory Workshop}, 2008,
  pp. 169--173.

\bibitem{Villard13}
J.~{Villard} and P.~{Piantanida}, ``Secure multiterminal source coding with
  side information at the eavesdropper,'' \emph{{IEEE} Trans. Inf. Theory},
  vol.~59, no.~6, pp. 3668--3692, 2013.

\bibitem{Schieler14}
C.~{Schieler} and P.~{Cuff}, ``Rate-distortion theory for secrecy systems,''
  \emph{{IEEE} Trans. Inf. Theory}, vol.~60, no.~12, pp. 7584--7605, 2014.

\bibitem{Kaspi15}
Y.~{Kaspi} and N.~{Merhav}, ``Zero-delay and causal secure source coding,''
  \emph{{IEEE} Trans. Inf. Theory}, vol.~61, no.~11, pp. 6238--6250, 2015.

\bibitem{Boyd04}
S.~Boyd and L.~Vandenberghe, \emph{Convex Optimization}.\hskip 1em plus 0.5em
  minus 0.4em\relax Cambridge university press, 2004.

\bibitem{Cover06}
T.~M. Cover and J.~A. Thomas, \emph{Elements of Information Theory},
  2nd~ed.\hskip 1em plus 0.5em minus 0.4em\relax New York: John Wiley {\&}
  Sons, 2006.

\bibitem{berger:1971}
T.~Berger, \emph{Rate Distortion Theory: A Mathematical Basis for Data
  Compression}.\hskip 1em plus 0.5em minus 0.4em\relax Englewood Cliffs, NJ:
  Prentice-Hall, 1971.

\bibitem{ElGamal11}
A.~El~Gamal and Y.-H. Kim, \emph{Network Information Theory}.\hskip 1em plus
  0.5em minus 0.4em\relax Cambridge University press, 2011.

\end{thebibliography}
\end{document}